\documentstyle[prl,aps]{revtex}

\input{epsf.sty}

\newcommand{\be}{\begin{equation}}
\newcommand{\ee}{\end{equation}}
\newcommand{\bea}{\begin{eqnarray}}

\newcommand{\eea}{\end{eqnarray}}
 
\newcommand{\ba}{\begin{array}}
\newcommand{\ea}{\end{array}}
\newcommand{\cao}{{\c{c}\~ao}}

\def\ls{\mathrel{\lower4pt\vbox{\lineskip=0pt\baselineskip=0pt
           \hbox{$<$}\hbox{$\sim$}}}}
\def\gs{\mathrel{\lower4pt\vbox{\lineskip=0pt\baselineskip=0pt
           \hbox{$>$}\hbox{$\sim$}}}}
\begin{document}

\twocolumn[\hsize\textwidth\columnwidth\hsize\csname
@twocolumnfalse\endcsname

\title{Charge Asymmetry in the Brane World and Formation of Charged Black Holes}

\author{Herman J. Mosquera Cuesta$^{1,2,3}$\footnote{herman@ictp.trieste.it}, 
Andr\'e Penna-Firme$^{1,4}$\footnote{apennafi@ictp.trieste.it}, 
Abdel P\'erez-Lorenzana$^{1,5}$\footnote{aplorenz@ictp.triste.it} }

\address{
$^1$The Abdus Salam International Centre for Theoretical Physics,
I-34100, Trieste, Italy \\ 
$^2$Centro Brasileiro de Pesquisas F\'{\i}sicas,
Laborat\'orio de Cosmologia e F\'{\i}sica Experimental de Altas Energias\\
Rua Dr.  Xavier Sigaud 150, 22290-180,  RJ, Brasil\\   
$^3$Centro Latinoamericano de F\'{\i}sica (CLAF),  
Avenida Wenceslau Braz 173,  Rio de Janeiro, RJ, Brasil\\ 
$^4$Universidade Federal do Rio de Janeiro (UFRJ), 
Faculdade de Educa\cao, Av. Pasteur, 250, 22290-180, 
RJ, Brasil\\ 
$^5$Departamento de F\'{\i}sica,
Centro de Investigaci\'on y de Estudios Avanzados del I.P.N.\\ 
Apdo. Post. 14-740, 07000, M\'exico, D.F., M\'exico }

\date{June, 2002}

\maketitle

\begin{abstract}
In theories with an infinite extra dimension, free particles localized on
the brane can leak out to the extra space. 
%%%%%%%%%%%%%%%%%%%%%
We argue that if there were color confinement in the bulk, electrons 
would be  more able to escape than quarks 
and than protons (which are composed states).  
%%%%%%%%%%%%%%%%%
Thus, this  process generates an electric charge asymmetry on brane matter
densities. A  primordial charge asymmetry during Big
Bang Nucleosynthesis era is  predicted.
We use current bounds on this and on electron disappearance 
to constrain the parameter space of these models.
Although the generated asymmetry is generically small, it could be
particularly  enhanced on large densities as in
astrophysical objects, like massive stars.
We suggest the possibility that such accumulation of  charge may be
linked, upon supernova collapse, to the formation of a charged Black Hole
and the generation of Gamma-Ray Bursts.
\end{abstract}

%\vskip1ex

\pacs{11.10.Kk; 04.50.+h; 98.80.-k; 98.70.Rz}
]

%%%%%%%%%%%%%%%%%%%%%%%%%%%%%%%%%%%%%%%%%%%%%%%%%%%%%%%%%%%%%%%%%%%%%%
%\section{Introduction}
%%%%%%%%%%%%%%%%%%%%%%%%%%%%%%

Extra dimensions could exist in nature and yet remain hidden to 
current  experiments. Motivated by the old ideas of Kaluza and Klein and
the modern string theory one has usually believed that such extra
dimensions should be compact. 
Recently, however, it has been  suggested  the
possibility that such extra dimensions  can be rather
infinite~\cite{lisa99a,kaloper,rubakov83} 
and still be hidden to our eyes. 
The explanation of why it is so comes from the concept of
localization of particles~\cite{rubakov83,likken,Jackiw,dvali} 
in a higher dimensional space (the bulk) 
around a fixed point that defines a
four dimensional hypersurface (the brane). 
Thus,  at low energy, observed particles  
would behave just as four dimensional fields. 
A simple realization of this scenario 
appears on five dimensional theories
where the background space is given
by the Randall-Sundrum (RS) metric~\cite{lisa99a}
 \be
 ds^2 = e^{-2k|y|} \eta_{\mu\nu} dx^\mu dx^\nu - dy^2~;
 \ee
where the parameter $k$ relates the fundamental five dimensional gravity
scale, $M_*$, with the now effective Planck scale, $M_P\sim 10^{19}$~GeV,
through  $M_*^3 = k~M_P^2$.
 
The above metric induces an effective potential for gravitons that has the
form of a volcano barrier~\cite{likken} around the brane 
(located at $y=0$).
The corresponding spectrum of gravitational perturbations 
has a massless bound state localized at the brane
and a continuum of bulk modes with suppressed couplings to brane fields.
The bound state is the true graviton that mediates the standard four
dimensional gravity interaction. Bulk modes exchange induces on the brane 
deviations from the four dimensional Newton's law of gravity
only at distances smaller than $k^{-1}$. Recent
experiments  that have tested gravity down to
millimeters~\cite{exp} require that  $k\gs 10^{-3}$~eV, 
which  means $M_*\gs 10^8$~GeV. 

Massless scalar fields on RS background have a similar spectrum, and so 
a bound state too~\cite{borut}. 
Localization of fermions also takes place provided one has a
domain wall on the background~\cite{rubakov83,Jackiw,rubakov2000a}.
Finally, localization of gauge particles is also 
possible~\cite{dvali,rubakov2000b}. 
%%%%%%%%%%%%%%%%%%%%%%%%%%
Last, however, according to Ref.~\cite{rubakov2000b} that we will follow heron, 
needs the existence of more extra flat dimensions 
compactified to small radii, $R\sim M_*^{-1}$.
In such a setup there is  a localized zero mode gauge boson, which 
alone satisfies the 4D Gauss law. 
Bulk modes introduce small corrections at short distances, though. 
The introduction of more compact dimensions, 
nevertheless, does not affect the model of localization of matter fields
since all heavy modes, associated to the compact space, 
shall always be integrated out from the theory. 
%Thus, the  model we are considering would be just an 
%effective zero mode approach itself.
%%%%%%%%%%%%%%%%%%%%%%%%%%%%%
%Hence, the so called brane world appears to be a viable possibility. 

True localization, however, takes place only for massless
fields~\cite{rubakov2000a,rubakov2000c}. 
In the  massive case the  bound state
becomes metastable and able to leak  into the  extra space. 
Such a process  appears as  
an apparent source for the violation of conservation
laws in the effective four dimensional world.
The associated escape rate  can be calculated in 
specific models for localization of free fields~\cite{rubakov2000a}.
Other interactions may also affect 
the escape rate by trying to keep the particles on the brane. 
So happens to
quarks, provided there is confinement in color interactions along all
dimensions~\cite{rubakov2000b}.

As we argue along this letter, in such a scenario, 
the overall effect would be  
the generation of a tiny electric charge asymmetry, 
due to the escaping  of electrons, 
which may be accumulated on large brane matter densities.
For instance, out of the matter asymmetry on the early Universe.
Current bounds on electron disappearance and on
charge asymmetry at the Big Bang Nucleosynthesis (BBN) 
era can then be used to  constrain the parameters of the model. 
We  suggest that this mechanism  could account 
for the formation of charged Black Holes (BH's), 
for instance after a supernova collapse. 
One possible implication of this would be 
the triggering of Gamma-Ray Bursts from vacuum polarization as proposed 
in Ref.~\cite{ruffini2001}.

%%%%%%%%%%%%%%%%%%%%%%%%%%%%
%\section
{\it Particle leakage and generation of charge asymmetry.--} 
In the models under consideration  massive  fields 
are actually  metastable states.
This happens because the  mass 
lifts the bound state on the spectrum, whereas  
the continuum of modes remains unaffected, and thus, 
our formerly localized
mode turns out to be  not longer  the lightest state.
Indeed, there would be a non zero probability for the
escaping of brane fields into the bulk~\cite{rubakov2000a,rubakov2000b}. 
One can check  this  by considering the localization mechanism
of a massive fermion. 
%%%%%%%%%%%%%%%%%
We stress that the following arguments do not preclude the
theory of having  small extra compact dimensions. 
So, our results are
consistent with the localization of gauge fields~\cite{rubakov2000b}.
To be clear, we open a parenthesis to 
remain the reader  how the localization of the photon works.
In  presence of $n$ extra flat dimensions,
with coordinates $\theta_i$, 
the action for a photon gauge field,  
truncated at the zero mode level of $\theta$, has  
the form: 
 \[ S_\gamma \approx
	-{1\over 4 } \int\! d^4x dy e^{-kn|y|}F_{\mu\nu} F^{\mu\nu} + 
	(\mbox{heavy modes})~.
  \]  
Hence the truely $y$-independent zero mode is normalizable, and thus it is
localized at the brane. For fermions the presence of $\theta$ dimensions
only rescale the normalization of the zero mode, so we can safely 
treat the problem as effectively five dimensional.
%%%%%%%%%%%%%%
Following Ref.~\cite{rubakov2000a},  
we consider a massive fermion, $\psi$, coupled to  
a domain wall background described by a scalar field, which 
in the thin wall approximation has the form $\phi(y)= v~sgn(y)$. 
The $\psi$ coupling to $\phi$  splits the chiral components 
of the four-component fermion  such that only one of them gets localized
around the wall.
Thus, in order to give rise to a four dimensional mass term we are forced
to introduce two five dimensional four-component fermions, $\psi_1$;
$\psi_2$, which will develop left and right localized modes,
respectively.  For simplicity one accommodates those fields in an
eight-component  array of the form $\Psi =(\psi_1,\psi_2)$.
With this notation the respective action can be written 
as~\cite{rubakov2000a}
 \be 
 S = \int\! d^4x dy \sqrt{g}~ \bar\Psi \left( i \gamma^a\nabla_a + 
 \phi\sigma_1 + \mu\sigma_2 \right) \Psi~;
 \ee
where  $\nabla_a$ is the spinor covariant
derivative with respect to the five dimensional metric $g_{ab}$
and $\sigma_i$ are the Pauli matrices 
acting on the $\psi_1$, $\psi_2$ space.
The general form of the Dirac equation 
in terms of chiral components
$\Psi_{L,R}$ is then given by 
\[
  e^{k|y|}\gamma^\mu p_\mu \Psi_{L,R} \pm \partial_5 \Psi_{R,L}- 
  \left(\phi(y) \sigma_1 + \mu\sigma_2  \right) \Psi_{R,L} = 0~.
\]
This  equation has a continuum of massive modes~\cite{rubakov2000a}  
that starts 
at $m=0$, for $m^2 = p_\mu p^\mu$, 
and a localized massive mode 
(that has radiation  boundary conditions at $y\rightarrow \pm \infty$) 
for which the mass eigenvalue is complex: 
$m = m_0 + i \Gamma$, where the physical mass $m_0 = (1-k/2\,v)\,\mu$.
The  imaginary part on $m$ 
reflects the metastable nature of the state. 
$\Gamma$ is then interpreted as the `life time' of the particle, from
the brane point of view.

In  the case where $m_0\ll k < v$  the probability for escaping goes as
 \bea
 \label{gamma}
 \Gamma = m_0 \cdot \left(\frac{m_0}{2k}\right)^{2v/k-1}
 \cdot\frac{\pi}{[\Gamma(v/k+\frac{1}{2})]^2},
 \eea 
This case   will probe  to be the most interesting one 
on our discussion below. 
By taking $k\gg m_e= 5.11\cdot 10^{-4}$~GeV, we state our interest 
to models  where $M_*> 10^{12}$~GeV.
The theory of escaping matter is thus described by three 
parameters: the mass of the particle, $m_0$; the vacuum parameter, $v$;
and the fundamental scale, $M_*$, encoded in the parameter~$k$.

{}For charged particles, their lost into the bulk 
account for an apparent violation of the conservation
of charge~\cite{rubakov2000b}, say for instance through the process 
$e^-\rightarrow nothing$. 
%%%%%%%%%%%%%%%%%%%%%%%%%% 
The picture  is  intrinsically  a higher dimensional problem, 
and so, four dimensional Gauss law is not longer valid, 
though causality is conserved.
In fact, a particle moving
with a trajectory perpendicular to the brane follows the world line: 
$y(t) = \ln(1 + k^2 t^2 )/2 k$.
It can be seen as a charged current which induces  an
electromagnetic field on the brane 
that dies away causally with time~\cite{rubakov2000b}. 
Indeed,  out of the
light cone ($|{\vec r}|>t$) the Coulomb potential of the charge remains:
$\phi({\vec r})\sim 1/r$, whereas deep 
in the light cone the electric field 
gradually disappears as $E\sim  r/t^3$.
This problem has a gravitational analogous 
studied in Refs.~\cite{rubakov2000c,giddings}.
%%%%%%%%%%%%%%%%%%%%%

While escaping away is certainly possible
for fundamental particles, it is  not so simple  for
composed states. These do not leak away from the brane 
as a whole so easily since
this needs the simultaneous leakage of all components, which
substantially reduces the escaping probability.  
%%%%%%%%%%%%%%%%%
For instance, for protons one naively gets a negligible escape rate: 
$\Gamma_P \sim m_P P_u P_u P_d\cdots$, 
with $P_{u,d}$ the escaping probability 
of quark  components.

In hadrons, confinement of color interactions may  prevent
the escape of a single colored component~\cite{rubakov2000b}.
%%%%%%%%%%%%%%%%%%%%%%%%%%%%%%%%%
It is believed that due to confinement 
the color flux does not spread around as gravity
and electromagnetism do. 
It rather forms flux tubes connecting quarks
that act  as an unidimensional strong attractive 
force that grows  with  distance, 
in spite of the orientation of quarks on space. 
It is then natural to expect that color will also be confined along the 
fifth dimension,
%%%%%%%%%%%%%%%%%%%%%%%%%%%%%%%%%%%%%
thus, making the trapping on the brane even stronger.
This also  forbids processes that may induce proton decay, and the sudden
creation of colored composed states from uncolored ones.
This asymmetric nature between electrons and protons 
translates into  the only  disappearance of electrons from the brane, 
and induces an  excess of charge (in the form of protons)
in any actual brane matter density (otherwise neutral).
%%%%%%%%%%%%%%%%%%%%
Notice, however, that if color were not confined on the bulk, 
a charge asymmetry will be generated any way
due to the  different charge (mass) of electrons and quarks, but 
we will not discuss this case in here.
%%%%%%%%%%%%%%%%
Next, let us  confront this result 
with some known current bounds that in turn will 
constrain the $v$-$M_*$ space parameter.

%section

{\it Bounding the parameter space.--} 
The first straightforward constraint comes from the bound on electron
disappearance, which is about
$\tau_{e\rightarrow nothing}> 4.2\cdot 10^{24}$~yr~\cite{edecay}. 
This implies that escape rate of electron should be 
 \be 
 \label{gexp}
 \Gamma \ls \Gamma_{exp} \equiv 1/ \tau_{e\rightarrow nothing}  = 
 4.96\cdot 10^{-57}~{\rm GeV}.
 \ee
A second phenomenological constraint comes from the fact that the fundamental
scale is supposed to be the largest scale on the theory, and thus $v\leq M_*$. 
In Fig.~1, we have plotted 
the curves that saturate both the  bounds. 
The allowed region in the parameter space for any viable model 
lies between both the curves and well within
the region where $k<v$. 
In fact,  the lower border
can be  described by a (phenomenological) 
power law relationship between $v$ and $M_*$ of the form:  
 $v \approx  (M_*/ M )^{1.91} M_* >  k$.
Hence, one concludes that all scenarios where $k\geq v$ are excluded on the
basis of current experimental bounds. 

Future improvements on present limits on electron disappearance would lift 
the lower border on Fig. 1. The effect, however, 
may be very small even for substantial improvements.  
{}For instance, an increment of 
nine orders of magnitude on the bound, that would render
$\tau_{e\rightarrow nothing}> 10^{33}$ yr,
would hardly change the allowed region in  parameter space 
as it is clear from Fig. 1, where the effect of this 
hypothetical bound is also depicted.

%\section

{\it Charge asymmetry at BBN era}.--
On the early Universe, a primordial charge asymmetry 
can be induced out of the 
primordial  matter-antimatter asymmetry.
The point is that in a balanced Universe, where 
electrons and positrons are equal in number, no charge asymmetry 
can be induced, since the escape rate has no dependence on the sign 
of the charge. However, since there is in fact an initial matter 
to antimatter  asymmetry per photon of order 
$\Delta \eta_m/ \eta_\gamma\sim 10^{-10}$,
some deficit of charge can indeed be generated. 
Though the  above  asymmetry is originally  in the 
form of baryons, some processes (as sphalerons) are expected to 
generate a similar lepton asymmetry. 
Thus, we would be looking into an initially neutral Universe where 
most of charged particles are equilibrated by an antiparticle, 
except by a very little amount of protons, as much as $\Delta \eta_m$,
that are compensated only by electrons.
Out of these initial $\Delta \eta_m$ free electrons, some will
leak out from the brane, such that at the BBN era 
the remaining  
$\eta^e_{BBN}= \Delta \eta_m\cdot \exp(-\Gamma~\tau_{BBN})$
electrons will not balance the charge equilibrium. The
density of charged particles in excess per photon would be:
 \[
 \Delta q_{BBN} = \left({\Delta \eta_m\over \eta_\gamma}\right)\,
 \left(1 -  e^{-\Gamma ~\tau_{BBN}}\right)\approx 
 10^{-10}\cdot \Gamma ~\tau_{BBN}~.
\]
A net charge density on early Universe would produce large scale relic
electric fields.
%%%%%%%%%%%%%%%%%%%%%%%
Certainly in a truly  homogeneous space 
the presence of a uniform charge density $\Delta q_{BBN}$ 
cannot give rise to any field. However, early Universe 
was not really homogeneous, 
otherwise not even structure formation could be possible. 
Indeed, there was a primordial inhomogeneity of order 
$\delta \rho/\rho\sim10^{-5}$. 
%The electric field would then
%have a coherent length probably larger that a horizon distance, 
%$|E|\ls l_h e \Delta q_{BBN}$. 
%%%%%%%%%%%%%%%%%%%%%%%%%
Such fields,  might be the
seeds for the  observed galactic  and extragalactic 
magnetic fields, but we will not address this possibility here.
These relic fields may also generate observable cosmic ray anisotropies.
Current limits impose the bound~\cite{orito} $\Delta q\ls 10^{-39}$.
This gives, however, a weaker bound for the escape rate than the one
already mentioned above. In fact one gets $\Gamma\ls 10^{-54}$~GeV 
($\tau_e\gs 10^{21}$~yrs). 
A stringent bound is obtained from the effect of a background charged
Universe on  primordial Helium production~\cite{masso}, which gives
$\Delta q\ls 10^{-43}$. That implies
 \be 
 \label{gbbn}
 \Gamma \ls \Gamma_{BBN} =
 6.58\cdot 10^{-58}~{\rm GeV}~.
 \ee
This also means an enhancement on the electron disappearance 
bound by a factor of seven.
The curve on the parameter space associated to the saturation of this
last bound  has been also plotted on Fig. 1, though it does not affect
the previous picture at all.

%\section

{\it Accumulation of charge on Stellar Objects.--} 
Though the charge asymmetry induced by the leakage of charge
might be too small  as to be directly detected
in terrestrial laboratories, it could be relevant in longevous
astrophysical objects which contain a large number of  particles.
In order to estimate this asymmetry for dense media
one  has to consider 
the presence of internal forces that keep those systems all
together in equilibrium in the empty interstellar space.
The attractive interaction that compensates the internal pressure  comes
mainly from the self-gravitational interaction  of the system. 
This attractive force will also tend to keep the particles on the brane, 
where the source is.
{}From the five dimensional point of view the effect of these 
internal forces  will be to locally lift  the trapping barrier,
thus, effectively reducing the escape rate.
%%%%%%%%%%%%%%%%%%%%%
One can account for this effect by introducing 
a phenomenological function,  $x(\vec r)$, 
to locally measure the  amount
of time along which the particle behaves as effectively free, 
and able to escape. 
Thus, one can writes  at zero order  the particle density 
as $\eta(\vec r,t) \approx \eta(\vec r) e^{-x(\vec r)\, \Gamma\, t}$.
{}From this equation one notices that the quantity $x\Gamma$ has the
physical meaning of being the effective escape rate, as modified by the
presence of interactions. So,  $x$ plays the role of a structure
factor that  encodes the internal forces of the medium  
at which the particle is subjected. 
Now, as the total number of remaining electrons at any given time
is given by 
 $N_e(t) = \int\!dV\, \eta(\vec r,t)$, and 
 due to the smallness of $\Gamma$, 
the induced charge asymmetry can be expressed in the form:
 \bea
 \Delta q(t) \equiv \frac{N_p - N_e(t)}{N_e(0)} =
 1- \frac{ N_e(t)}{N_e(0)} \approx \bar{x}~\Gamma~t~; 
 \label{carga}
 \eea 
where $\bar x$ indicates the volume average of $x$ that becomes  
the actual  phenomenological parameter, yet to
be calculated for realistic cases.
%%%%%%%%%%%%%%%%%%%%%%%
It is expected that for most (not so dense) media $\bar x\approx 1$, 
however, it may be important for tightly bound systems as neutron stars.
%%%%%%%%%%%%%%%%%%%%%%%

Let us now consider a typical star to estimate how much of asymmetry 
it can accumulate along its  life, $t_*$, 
which may be almost as large as the age of the universe. 
Thus, taking $t_*=10^{10}$~yr into Eq.~(\ref{carga}), 
combined  with the bounds on Eqs.~(\ref{gexp}) and (\ref{gbbn})
we get 
 \be
 \label{dq}
 \Delta q(t_*)\ls  10^{-14,~-16}~.
 \ee

{\it Charged Black Holes and Gamma-Ray Bursts--}
A tiny charge asymmetry would  generate small electromagnetic fields
which  can be substantially magnified~\cite{SNcollapse} by some
astrophysical processes such as  supernova core collapse. The presence of
a non zero total  charge on stars progenitor of supernovae could give 
rise to other
fundamental  phenomena like the ulterior formation of charged Black
Holes. The origin of charged BH's has  been a long standing puzzle in
relativistic astrophysics after the  derivation of Reissner-Nordstr\"om
solution of Einstein's equations.  Thence, this brane world inspired
charge-leaking mechanism provides a natural  explanation for their origin
upon stellar evolution of massive stars, thus,  solving  such an enigma.

An interesting consequence of all this could be the generation of 
Gamma-Ray Bursts (GRBs) as already suggested in Ref.~\cite{ruffini2001}. 
However, the open point is whether particle leakage can account for the
required charge asymmetry to switch on such a  process.   
The answer to this question appears to be on the positive. According to
Ref.~\cite{ruffini2001}, vacuum polarization   occurring during the
formation phase of  a  Reissner-Nordstr\"om Black Hole may generate
GRBs, provided that the net charge-to-mass ratio, 
$\xi = Q/(M\sqrt{G})$; be of the order $0.01$ to one. 
The limit $\xi=1$ corresponds to a critical charged Black Hole. 
Nevertheless, it is likely that even smaller values of $\xi$
could still render workable the quoted mechanism. 
In any case, the above condition translates 
into an effective charge asymmetry 
per nucleon in the collapsing matter:  
$\Delta q_{BH} \approx \xi~ 10^{-18}$.  
This value is at least two orders of magnitude smaller than the upper
bound on Eq.~(\ref{dq}), and thus, it can indeed be provided by the
charge leakage.  
Moreover, for generating such an asymmetry,  one
concludes that the electron has to disappear from the brane on a  time,
$\tau_e$, within the range
 \be 
 4.2\cdot 10^{24}~{\rm yr} \ls \tau_e \ls 2.08 
 ~\xi^{-1}\cdot 10^{28}~{\rm yr}~.
 \label{tgrb}
 \ee
The allowed region in the parameter space that is  consistent 
with these bounds is yet very narrow, 
thus, pointing towards a very specific class of models.
Indeed, even 
for a $\xi\sim 10^{-5}$ that gives $\tau_e \ls 10^{33}$~yr;  one gets the
same narrow strip  already depicted in Fig. 1 within current and
hypothetical future limits.

%\section{Conclusions} 

{\it Summarizing--} In brane world models  massive fundamental  states
become metastable with a non zero probability for escaping away into the
bulk. Nevertheless, confinement of color interactions will prevent  the
leakage of quarks, 
while protons  have a negligible scape rate.
Therefore, only electrons would be effectively
missed,  and so, unbalancing  the neutralness of brane matter densities. 
As a result modest charge asymmetries can be generated on the 
early Universe and in stellar objects, which may  have  observable
effects.
We confronted this result with current bounds on  electron
disappearance and  relic electric charge at BBN era,  and consequently
constrained the parameters of the brane world model. 
We have suggested the possibility that  supernova collapse of massive
stars may indeed form charged Black Holes.   
We shown that the  charge of such a  Black Hole can 
be of the right order as to trigger GRBs via vacuum polarization. 
There is a non zero parameter space
consistent with this picture that also points towards an electron
disappearance rate non far from current experimental limits.

%{\it Acknowledgements.}   
%%%%%%%%%%%%%%%%%%%%%%%%%%%%%%
We thank A. Mazumdar for useful conversations. The work of APF  was supported
by a grant of Conselho Nacional de Desenvolvimento Cient\'\i fico e
Tecnol\'ogico, CNPq--Brasil. HJMC thanks Funda\c c\~ao de  Amparo \`a Pesquisa
do Estado do Rio de Janeiro (FAPERJ) for a grant-in-aid.

%%%%%%%%%%%%%%%%%%%%
\begin{figure}%[ht]
%\vskip5em
\centerline{
\epsfysize=120pt
\epsfbox{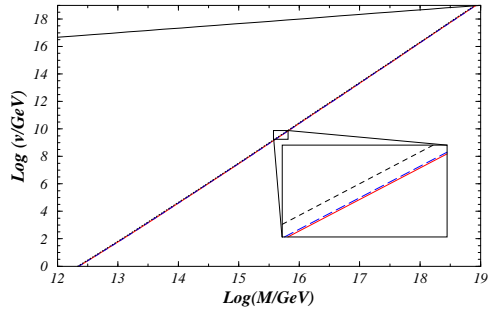}
}
\vskip1em

\caption{Allowed parameter space for brane world models. Upper line stands for
$v=M_*$. Lower line is the limit imposed by $\Gamma_{exp}$
(continuous line) and $\Gamma_{BBN}$ (dashed line). 
Dotted line on the blown up zone pictures an hypothetical future limit 
on electron disappearance of about $10^{33}$~yr.}
\end{figure}
%%%%%%%%%%%%%%%%%%
\vskip2em

\end{document}